\documentclass[newstyle,twocolumn,proceedings]{rmaa}
\usepackage{epsfig}

\def\apj{ApJ}
\def\mnras{MNRAS}
\def\aj{AJ}
\def\aap{A\&A}
\def\apjs{ApJS}

\renewcommand{\P}[1]{%
\ifnum#1=1\hbox{OW~168--326E}\fi
\ifnum#1=2\hbox{OW~167--317}\fi
\ifnum#1=3\hbox{OW~163--317}\fi
\ifnum#1=5\hbox{OW~158--323}\fi
\ifnum#1=0\hbox{OW~171--334}\fi}
\makeatletter
\title{Stellar Populations in Circumnuclear Star Forming Regions}
\author{ A. I. D{\'\i}az\altaffilmark{}, M. {\'A}lvarez-{\'A}lvarez\altaffilmark{} and M. Castellanos\altaffilmark{} 
  \affil{Universidad Aut{\'o}noma de Madrid, Spain}}

\fulladdresses{
\item  A. I. D{\'\i}az,  M. {\'A}lvarez-{\'A}lvarez and M. Castellanos: Depto. F{\'\i}sica 
Te{\'o}rica, C-XI,
  Universidad Aut{\'o}noma de Madrid, 28049 Madrid,
  Espa{\~n}a (angeles.diaz@uam.es, malvar4@roble.pntic.mec.es, marcelo.castellanos@uam.es).
}
\shortauthor{D{\'\i}az et al.}
\shorttitle{Circumnuclear Star Forming Regions}
\keywords{ISM: \ion{H}{2} regions  --- Stars: ionizing populations}

\abstract{%
We present a study of the stellar populations and gas physical conditions
in Circumnuclear Star Forming Regions (CNSFR) based on
broad  and narrow band photometry and spectrophotometric data, which
have been analyzed with the use of evolutionary population synthesis
and photoionization models. It is found that most CNSFR
show composite stellar populations of slightly different ages. They seem to
have the highest abundances in HII region-like objects, showing
also N/O overabundances and S/O underabundances by a factor of about
three. Also, CNSFR as a class, segregate from the disk HII region
family, clustering around smaller
$\eta$' values, and thereforefore higher ionizing temperatures. } 
\resumen{%
Presentamos un estudio de las poblaciones estelares y las condiciones
f{\'\i}sicas del gas en Regiones Circunnucleares de Formaci{\'o}n Estelar
(CNSFR) basado en fotometr{\'\i}a de banda ancha y estrecha y datos 
espectrofotom{\'e}tricos, que se han analizado mediante el uso de modelos
evolutivos de s{\'\i}ntesis de poblaciones y de fotoionizaci{\'o}n. 
Hemos encontrado que las CNSFR muestran
poblaciones estelares compuestas, de edades ligeramente diferentes. 
Parecen tener 
las abundancias m{\'a}s altas dentro de los objectos de tipo region
HII, a la vez que muestran tambi{\'e}n cocientes de N/O y S/O por encima y
por debajo de los valores solares en un factor de aproximadamente
3. Tambi{\'e}n, las CNSFR como clase, se segregan de la familia de regiones
HII de disco, agrup{\'a}ndose alrededor de valores de $\eta$' menores, y por lo
tanto temperaturas ionizantes m{\'a}s altas.}

\listofauthors{A.~I.~D{\'\i}az, M.~{\'A}lvarez-{\'A}lvarez, M. Castellanos}
\indexauthor{D{\'\i}az, A.~I.}
\indexauthor{{\'A}lvarez-{\'A}lvarez, M.}
\indexauthor{Castellanos, M.}

\begin{document}

%% This command is necessary to typeset the title, abstract, etc. 
\maketitle

%%
%% And here starts the text....
%%
\section{Introduction}
\label{sec:intro}
The inner parts of some spiral galaxies show higher star formation
rates than usual and this star formation is frequently arranged in a
ring or pseudo-ring pattern around their nuclei. This fact seems to 
correlate with the presence of bars and, in fact, computer models which
simulate the behavior of gas in galactic potentials have shown that
nuclear rings may appear as a consequence of matter infall owing to
resonances present at the bar edges (Combes \& Gerin 1985; Athanassoula 1992).

In general, Circumnuclear Star
Forming Regions (CNSFR), also referred to as ``hotspots'', are alike
luminous and large disk HII regions, but look more compact ad show
higher peak surphace brigtness  (Kennicut et al. 1989). In many cases
they contribute substantially to the emission of the entire nuclear
region. 

Their large H$\alpha$ luminosities, typically higher than 10$^{39}$ erg
s$^{-1}$ points to relatively massive star clusters as their ionization
source, which minimizes the uncertainties due to small number
statistics when applying population synthesis techniques (see
e.g. Cervi{\~n}o et al. 2002). These regions then constitute excellent
places to study how star formation proceeds in circumnuclear environments.

To this aim, we have combined broad-band photometry, narrow-band Balmer
emission imaging and spectrophotometric data for a sample of CNSFR
whose analysis can provide detailed information about both ionizing and
non-ionizing stellar populations.

%%%%%%%%%%%%%%%%%%%%%%%%%%%%%%%%%%%%%%%%%%%%%%%%%%%%%%%%%%%%%%%%%%%%%%%%%%%%%

\begin{figure*}
\setcounter{figure}{0}
\begin{minipage}[l]{17cm}
 \centering
 \psfig{figure=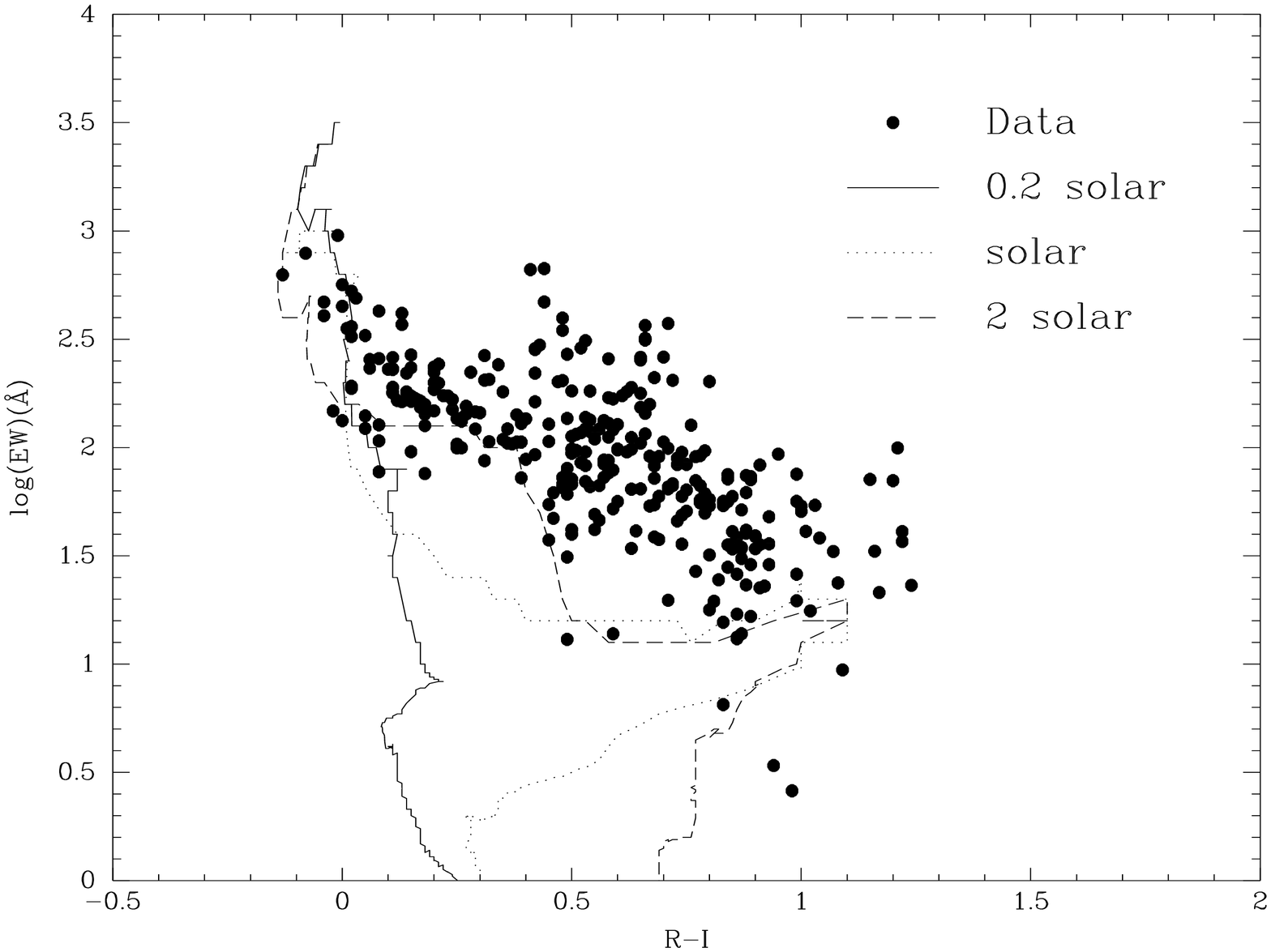,width=5.5cm,angle=270,clip=}
%\end{minipage}
%\begin{minipage}[r]{10cm}
 %\centering
\hspace{1.3cm}
 \psfig{figure=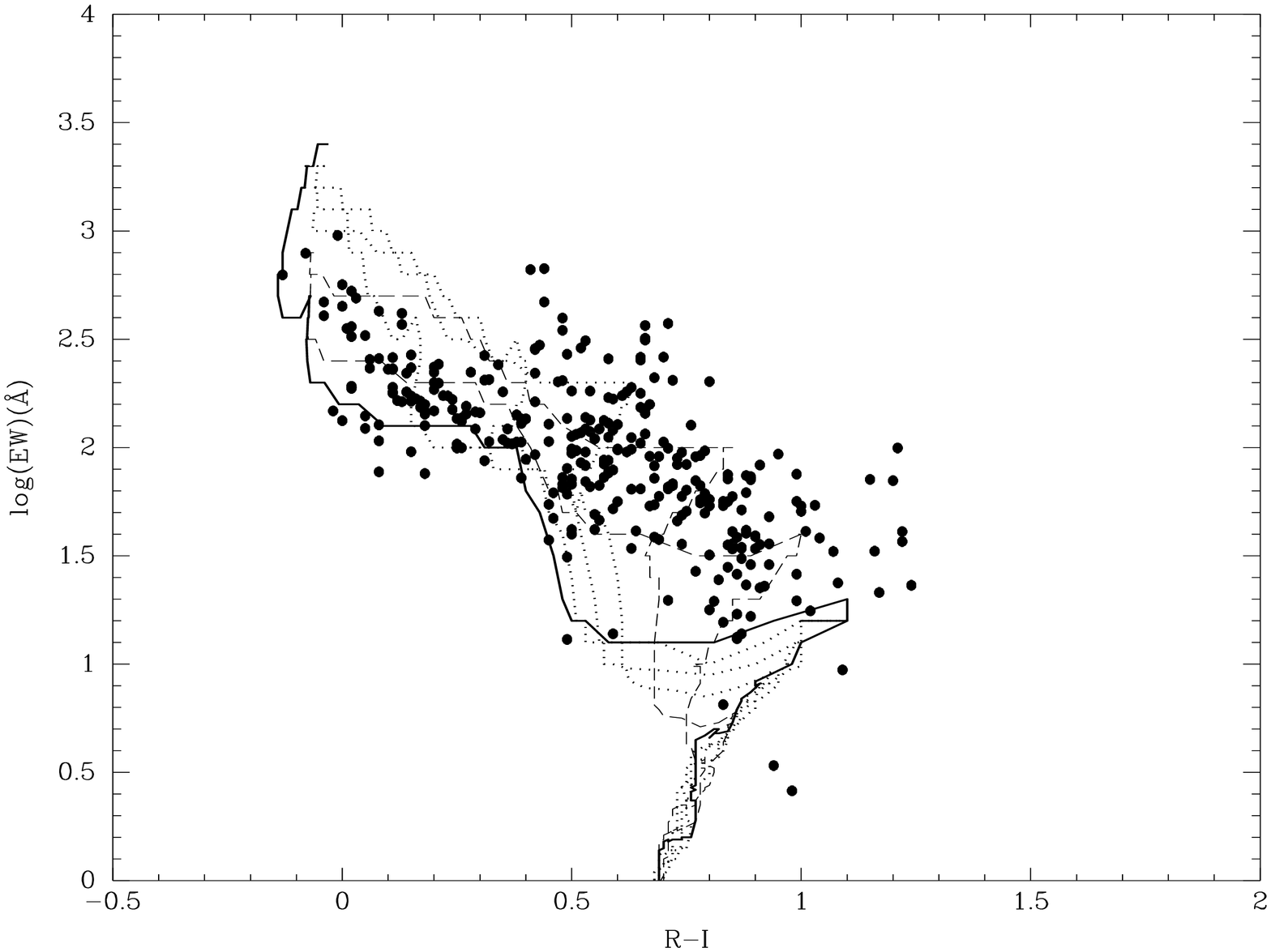,width=5.5cm,angle=270,clip=}
\end{minipage}
 \caption{Relative intensity (left) and equivalent width (right) of the
   WR blue `bump' {\it versus}  H$\beta$ equivalent width for SV98 models of
   three different metallicities as labelled. The data are shown as
   solid and open symbols as explained in the text.}
\end{figure*}
%%%%%%%%%%%%%%%%%%%%%%%%%%%%%%%%%%%%%%%%%%%%%%%%%%%%%%%%%%%%%%%%%%%%%%%%%%%%%

\section{Observations and Data Sample} 
\label{sec:obs}

\subsection{Photometric Data}
\label{subsec:phot}

Two data sets of broad and narrow band photometry were obtained. The
first one with the JKT telescope in the 
Observatory of the Roque de los Muchachos (ORM, La Palma, Spain), in the
V,R,I and H$\alpha$ filters included four galaxies. The details concerning
this first set of data can be found in D{\'\i}az et al. (2000). 
In the second one, with
the 1.52 m Spanish Telescope in Calar Alto Observatory (Almer{\'\i}a, Spain),
the sample was enlarged to 20 galaxies and the filters used were
U,B,R,I, H$\alpha$ and H$\beta$.
Spatial resolution is about 0.33 arcsec/pix. 
%Figure 1 shows images of
%one of the sample galaxies, NGC~4314, in the H$\alpha$ and U filters.

The observed galaxies are bright (14.1 $>$ B $>$ 9.6) and nearby (8.4
$<$ D (Mpc) $<$ 37.2) yielding liner scales between 41 and 181
pc/arcsec. They are spirals of different morphological type. 70\% of
them are barred (20\% strongly and 50\% weakly) and 25\% are unbarred. No
information is available for the remaining 5\%. Regarding interaction,
60\% of them have a neighboring galaxy closer than 4 arcmin and 40\% do
not. Finally,  65\% of the sample harbour an active nucleus: 10\% Sey1,
40\% Sey2, 15\% LINER, 30\% shows HII region nuclei and 5\% do not show any
signs of nuclear activity.

A total of 332 CNSFR in the sample galaxies have been measured and
analyzed. 

\subsection{Spectroscopic Data}
\label{subsec:spec}

Regarding spectrophotometry, we have analyzed data of moderate
resolution (1.4 - 2.5 {\AA}/pix) and wide coverage (3500 to 9700 {\AA}) obtained
with the INT and WHT telescopes (ORM). More details on these
observations can be found in P{\'e}rez-Olea (1996) and {\'A}lvarez-{\'A}lvarez 
et al. (2001). These configurations allow the observation of both the bright
oxygen ([OII] and [OIII]) and sulphur ([SII] and [SIII]) emission lines
necessary for the diagnostics of the emitting gas.

\section{Analysis of Broad-Band Colors: Stellar Population Synthesis Models}
\label{sec:colors}
The broad band colors have been analyzed in combination with the H$\alpha$
equivalent widths obtained through narrow filters, with the use of the
stellar population synthesis models by Leitherer et al (1999;
STB99). These models give predictions for  stellar populations of
different metallicities (between 0.05 and 2 times solar) and initial
mass functions (power laws with exponent $\alpha$= 2.35, 3.00, upper mass
limit m$_{up}$ = 100, 30 M$_{\odot}$ and lower mass limit  m$_{low}$ = 1 M$_{\odot}$).
The observed colors have been corrected for reddening using the
observed H$\alpha$/H$\beta$ ratio. Both single burst and continuous star
formation are considered.

%%%%%%%%%%%%%%%%%%%%%%%%%%%%%%%%%%%%%%%%%%%%%%%%%%%%%%%%%%%%%%%%%%%%%%%%%%%%%
\begin{figure*}
\setcounter{figure}{1}
\begin{minipage}[l]{16cm}
 \centering
 \psfig{figure=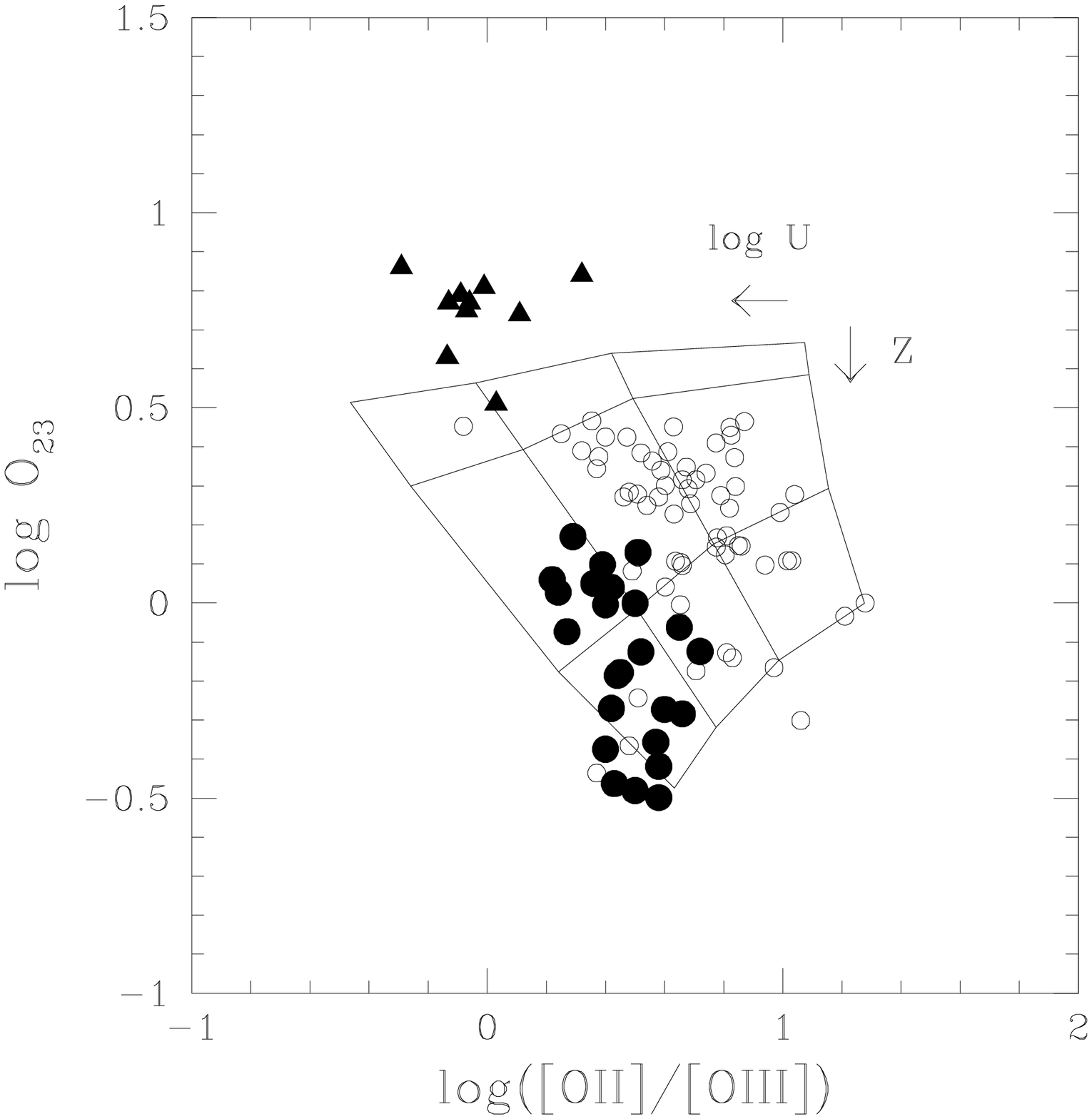,height=6.0cm,width=6.85cm,clip=}
%\end{minipage}
%\begin{minipage}[r]{10cm}
 %\centering
\hspace{1cm}
 \psfig{figure=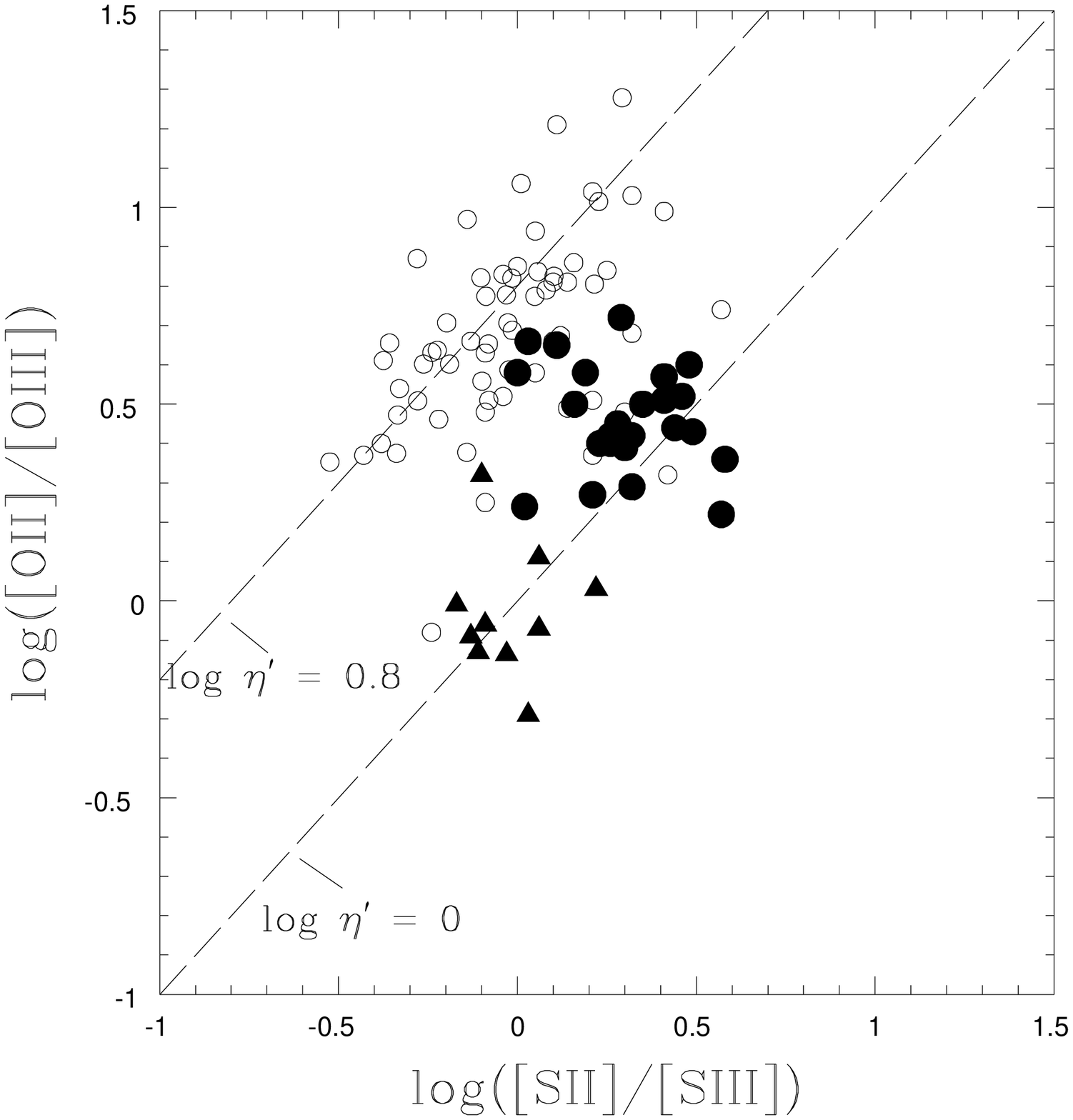,height=6.0cm,width=6.85cm,clip=}
\end{minipage}
 \caption{The logO$_{23}$ {\it vs} log([OII]/[OIII]) (left panel) and 
   log([OII]/[OIII]) {\it vs} log([SII]/[SIII]) (right panel) diagnostic
   diagrams. Solid symbols correspond to CNSFR of low (triangles) and
   high (circles) metallicity. Open symbols are high metallicity disk
   HII regions from the samples of Bresolin et al. (1999) and D{\'\i}az \& 
   P{\'e}rez-Montero 2000. Continuous lines show the predictions of
   photo-ionization models as explained in the text.}
\end{figure*}
%%%%%%%%%%%%%%%%%%%%%%%%%%%%%%%%%%%%%%%%%%%%%%%%%%%%%%%%%%%%%%%%%%%%%%%%%%%%%

In our analysis we have assumed that the H$\alpha$ emission comes from the
ionizing stars while the contribution of any underlying
stellar population should be detectable as emission in continuum
light, mostly at the longer wavelengths (redder colors). Figure 1
(left panel) shows
the equivalent width of H$\alpha$, EW(H$\alpha$), {\it versus} the (R-I) colour for
the observed regions. Single burst stellar populations for a Salpeter
initial mass function and different metallicities: 0.2 solar, solar and
twice solar are overimposed. It can be seen that only a small fraction
of the observed
objects (the very blue ones with high values of EW($H\alpha$) and the very
red ones with low values of EW($H\alpha$)) can be reproduced by the
models. Continuous star formation provides even  worse results. The
great majority of the regions can be explained only by the combination
of at least two populations of different ages. In order to explore this
possibility, we have
constructed models of two populations of the same metallicity and
slightly different ages, combined according to two parameters: t, which
represents the difference in age of the two assumed populations, and f,
which represents the ratio between the mass of the youngest population
and the total mass of the region. Models of this kind are shown in
Figure 1 (right panel) to be able to reproduce the observations in the 
EW($H\alpha$)- (R-I) plot. This situation looks however more complicated
when the (U-B)
colors are considered in the analysis. Not even the composite
population models are able to reproduce the location of the HII regions
in the (U-B)-(R-I) diagram, with about half of the regions showing (U-B)
colors much redder than any model prediction. This fact needs to be
studied further both from the observational and theoretical sides in
order to understand how star formation takes place in these HII regions.

%%%%%%%%%%%%%%%%%%%%%%%%%%%%%%%%%%%%%%%%%%%%%%%%%%%%%%%%%%%%%%%%%%%%%%%%%%%%%

\begin{figure*}
\setcounter{figure}{2}
\begin{minipage}[l]{16
cm}
\centering
 \psfig{figure=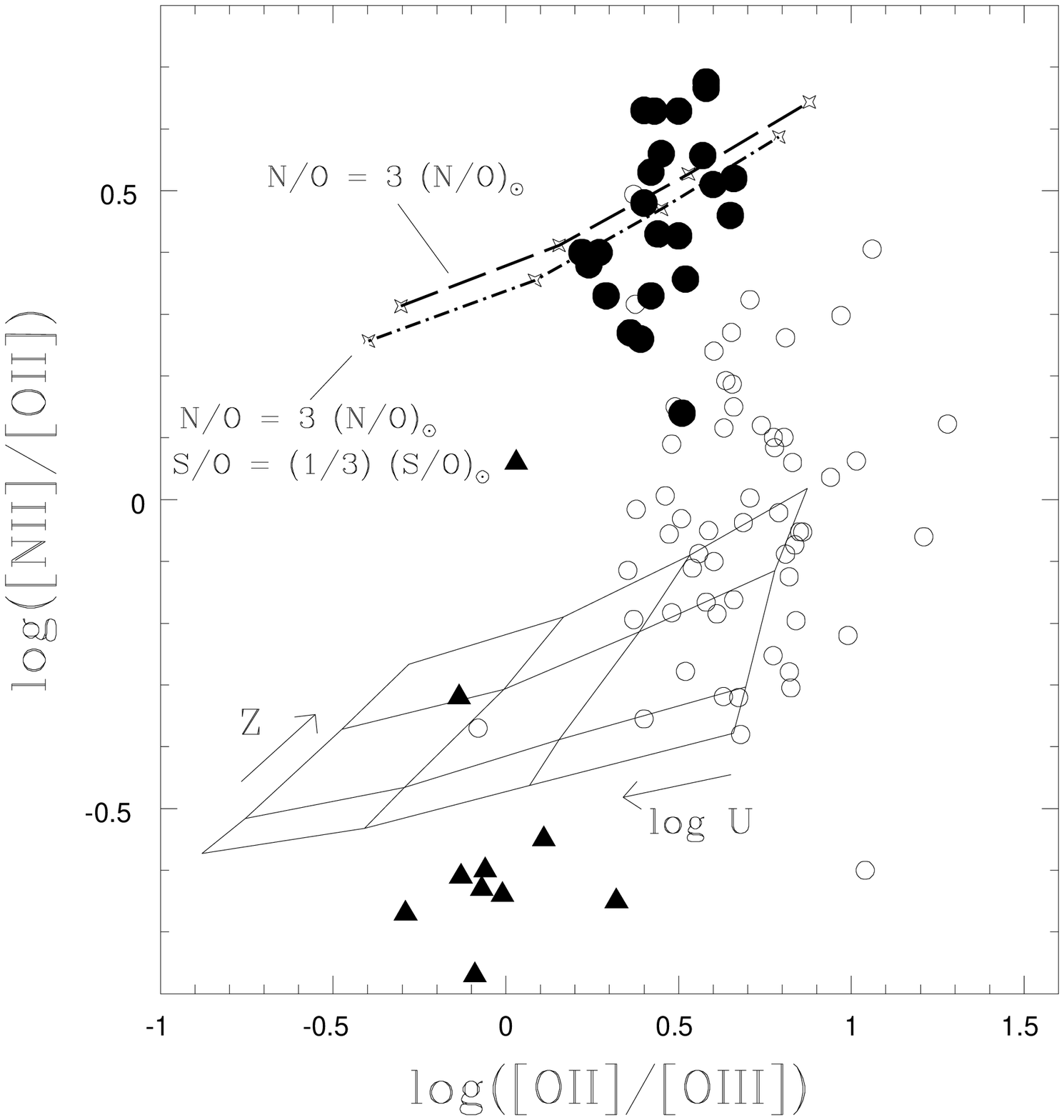,height=6.0cm,width=6.85cm,clip=}
\hspace{1cm}
%\end{minipage}
%\begin{minipage}[r]{10cm}
 %\centering
 \psfig{figure=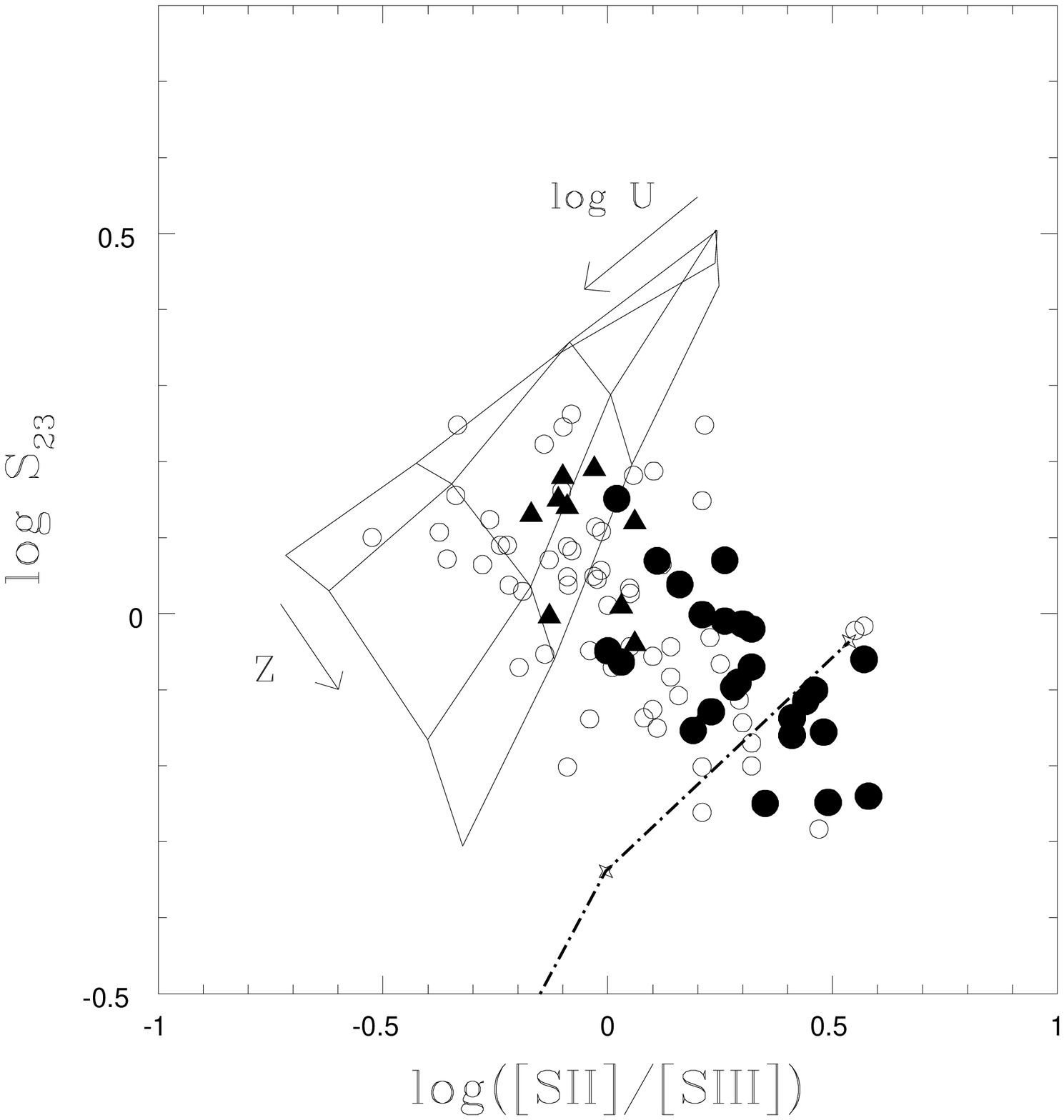,height=6.0cm,width=6.85cm,clip=}
\end{minipage}
 \caption{The log([NII]/[OII] {\it vs} log([OII]/[OIII]) (left panel) and 
   logS$_{23}$ {\it vs} log([SII]/[SIII]) (right panel) diagnostic
   diagrams. Symbols are as in the previous figure.}
\end{figure*}

%%%%%%%%%%%%%%%%%%%%%%%%%%%%%%%%%%%%%%%%%%%%%%%%%%%%%%%%%%%%%%%%%%%%%%%%%%%%%%%

\section{Analysis of Emission Line Spectra: Photo-ionization Models}
\label{sec:els}
The emission line data have been analyzed with the use of
photo-ionization models (CLOUDY; Ferland 1999). We have modelled the
HII regions using the simplest hypotheses: a single ionizing star,
spherical symmetry, uniform chemical composition and constant particle 
density. Any effects due to the presence of dust have not been
considered, except for the depletion of refractory elements by a factor
of ten. We have assumed the emission line spectrum of the HII region to
be controlled by the spectral energy distribution of the ionizing star,
which has been represented by Mihalas' atmosphere models of different
effective temperatures, the geometrical factors, represented by the
ionization parameter, U, and the chemical composition of the gas,
represented by its oxygen abundance relative to hydrogen, O/H. The
comparison of model predictions and data has been made with the use of
diagnostic diagrams.

Figure 2 (left panel) shows the abundance parameter log O$_{23}$=
([OII]+[OIII])/H$\beta$ {\it versus} the ionization parameter indicator 
log([OII]/[OIII]). Observed CNSFR are represented as solid symbols,
triangles for regions in NGC~3310 and NGC~7714 known to have a subsolar
oxygen abundance (Pastoriza et al. 1993, Gonz{\'a}lez Delgado et al. 1995),
and circles for the rest of the analyzed galaxies: NGC~1068, NGC~2903,
NGC~3351, NGC~3504 and NGC~5953. These latter regions have solar or
oversolar metallicity according
to the empirical criterion of D{\'\i}az \& P{\'e}rez-Montero (2000), {\it i.e.} O$_{23} \leq$
0.47 and -0.5 $\leq$S$_{23} \leq$ 0.28\begin{footnote}{The sulphur abundance parameter
  S$_{23}$ is defined as logS$_{23}$=
([SII]+[SIII])/H$\beta$}\end{footnote}.
As a comparison sample, open circles
represent disk HII regions from the samples of D{\'\i}az \&
P{\'e}rez-Montero (2000) and Bresolin et al. (1999) which meet the same criterion. The lines in the
diagram correspond to photoionization models of the same effective
temperature (37000 K) and different ionization parameter (from -2.0 to
-3.5) and O/H abundance (from 0.7 to 3 times solar). It can be seen
that most high metallicity CNSFR have O/H abundances higher in average
than their disk counterparts although they show comparable
([OII]/[OIII]) ratios. CNSFR also segregate from the disk HII regions in the 
([OII]/[OIII]) {\it versus} ([SII]/[SIII]) diagram (Figure 2, right panel). The
former cluster around the value of log$\eta$' = 0.0 while the latter do
around  log$\eta$' = 0.8. This parameter, $\eta$' is a measure of the hardness
of the ionizing radiation (see V{\'\i}lchez \& Pagel 1988) and seems to imply
higher ionizing temperatures for the CNSFR.

Regarding relative abundances, some peculiarities are observed
concerning both nitrogen and sulphur. These can be visualized in Figure
3. In the left panel we can see the location of the CNSFR in the
log([NII]/[OII] {\it versus} log([OII]/[OIII]) diagram. The high
metallicity CNSFR look clearly overabundant in nitrogen in comparison
to disk HII region, which is compatible with their apparent higher
metallicity.
The continuous lines in the diagram show the same photoionization
models as in Figure 2. For these models the N/O ratio has been kept to
its solar value. The dashed line, however, shows models with O/H and
N/O 3 times their solar values. The right panel shows the position of the
CNSFR in the logS$_{23}$ {\it versus} log([SII]/[SIII]) diagram. A
trend of decreasing S$_{23}$ and increasing [SII]/[SIII] seems to be
defined by the data points, which could be interpreted, in principle, 
as an increase
in metallicity and a corresponding decrease in ionization
parameter. This trend however cannot be reproduced by the models with
solar relative abundances which are shown, again, as continuous
lines. The only way to reproduce the location of the CNSFR implies a
reduction of the S/O abundance by a factor of about 3, as shown by the
dashed-dotted line. This reduction
does not affect the  [NII]/[OII] and [OII]/[OIII] ratios as can be seen
in the left panel.

%\section{Summary and Conclusions}
%\label{summ}

\acknowledgements This work has been partially supported by DGICYT
proyect AYA-2000-0973.

%% When using the rmaacite package, the \bibitem command should be of
%% the format: 
%%
%%             \bibitem[AUTHOR<YEAR>]{KEY} 
%%
%% so that the \cite{KEY}, etc. commands will work properly. 
%% 
%% If you are doing the citations manually, then you can just use
%% `\bibitem{}' instead. This will give you a warning about
%% `multiply-defined labels' which you can safely ignore.
%% 
\adjustfinalcols

\end{document}